# Transition Matrix Cluster Algorithms

David Yevick, Yong Hwan Lee
Department of Physics
University of Waterloo
200 University Ave. W.
Waterloo, ON N2L 3G1, Canada

We demonstrate that a series of procedures for increasing the efficiency of transition matrix calculations can be realized by integrating the standard single-spin flip transition matrix method with global cluster flipping techniques. Our calculations employ a simple and accurate method based on detailed balance for computing the density of states from the Ising model transition matrix.

Monte-Carlo Methods, Statistical Physics, Transition Matrix Methods, Phase Transitions, Cluster Algorithms

## 1 Introduction

Often the statistical properties of a set of global variables $\vec{E}(\vec{\alpha})$ that depend on an ensemble of stochastically varying local quantities $\vec{\alpha}$ must be determined. If the physically interesting regions of the global variables correspond to rarely occurring configurations of the local parameter space, the computational efficiency can generally be greatly improved through biased sampling. In particular, specializing to a single system variable, $E$, a proposed realization of a Markov chain, $\vec{\alpha}_{new}$, is related to the current configuration by a small random change, $\delta\vec{\alpha}$ according to $\vec{\alpha}_{new} = \vec{\alpha}_{current} + \delta\vec{\alpha}$ which leads to a change in the global variable from histogram bin $i$ corresponding to an average energy $E_i$ to a histogram bin $j$ with average energy $E_j$ ($i$ and $j$ can be identical). This proposal is then accepted in accordance with a rule that preferentially samples the physically relevant ranges of $\vec{E}(\vec{\alpha})$, in which case $\alpha_{current}$ is replaced by $\alpha_{new}$. After a sufficient number of steps, the bias introduced by the acceptance rule is appropriately removed. Biased sampling procedures such as the multicanonical [1] [2] [3] [4] [5] [6] [7] [8] [9] [10] [11] and Wang-Landau [12] [13] [14] techniques are particularly effective, enabling order-of-magnitude reductions in computation time without prior knowledge of system properties.

Markov chain procedures can be further extended by constructing a transition matrix $T$

such that for every transition from $E_i$ to $E_j$ before the acceptance rule is applied the element $T_{ij}$ is incremented by unity. After all transitions are recorded and each row in T is normalized, $T_{ij}$, coincides with the probability that the Markov chain evolves from a histogram bin $E_i$ transitions to a bin $E_j$ in a single unbiased Markov step, $\delta\vec{\alpha}$. [15] [16] [17] [18] [19] [20] [21] [22] [23] [24] [25] [26] The normalized eigenvector of $T$ with unit eigenvalue, typically obtained by repeatedly multiplying an initially random vector by $T$, corresponds to the density of states (infinite temperature probability distribution) $p\left(\vec{E}(\vec{\alpha})\right)$ associated with $\vec{E}(\vec{\alpha})$. [27] Since the transition matrix method retains all accepted and rejected transitions, it displays improved accuracy and scaling properties relative to the corresponding biased sampling technique as has been exhaustively demonstrated in previous articles, e.g. [28].

For the Ising model, $\delta\vec{\alpha}$ is generated by flipping a single, randomly selected spin. Since the transition matrix elements are formally independent of the acceptance rule, simple algorithms based on e.g. transition probabilities between microscopic states [16] [17], the ratio of transition matrix elements [29] [30] [17], and the exclusion of transitions to bins that have been previously visited a larger number of times [23] [27] [31] can be employed to generate $T$. In reality, however, such procedures do not provide sufficient accuracy when applied to e.g. the specific heat of systems near phase transitions because of the pronounced expansion in the number of accessible macrostates (with differing physical properties), which can only be sampled after numerous Markov steps. [15] Rather, accumulating transition matrix elements while annealing through a series of canonical, Metropolis [32] calculations at inverse temperatures that vary slowly from e.g. a large initial value to near zero such that $\beta \equiv 1/T = f(m)$ in the m:th computation step are far more precise than quasi-microcanonical methods that sample $\vec{E}(\vec{\alpha})$ within a narrow but similarly shifting region. [33] [34] [35] Here $f(m)$ is a monotonic function (the inverse temperature schedule) that smoothly interpolates between the initial and final inverse temperatures. The numerical accuracy is greatly improved if $f(m)$ varies slowly with $m$ near the inverse critical temperature. Various techniques have accordingly been advanced for optimizing transition matrix calculations. These include accumulating transitions from multiple independent Markov chains and dynamically adapting the temperature schedule to variations in the size of the accessible state space [15] by monitoring either the correlation time, [28] the convergence of the normalized histogram of samples as a function of magnetization [35] or the canonical entropy of either the full phase space or the phase space in the magnetization-energy diagram. [36] Additionally, the procedure can be accelerated through a renormalization strategy that generates an approximate density of states from the transition matrices of smaller subsystems. [28] [35] [37]

This paper advances an alternate and, to our knowledge, novel approach to increasing the accuracy of transition matrix methods by integrating cluster flips with single spin-flips. Although specialized to the Ising model and the Wolff algorithm, the method clearly should generalize to more complex systems for which the Wolff procedure can be replaced by machine learning approaches. [38] [39] In particular, while $T_{ij}$ is assembled only from transitions obtained by inverting a single, randomly selected single spin, global cluster flips generated by the Wolff method [40] (or any analogous procedure) are interspersed with certain of these transitions. While the Wolff procedure is considerably less computationally efficient than single spin flip, the integrated procedure if judiciously implemented attains a significantly increased level of accuracy, especially for large systems. This follows from the nearly random sampling of the accessible phase space by the Wolff algorithm, which contrasts with the slow diffusion of a single flip Markov chain. Therefore, following cluster flips with the single spin flips required to construct a transition matrix ensures that the sampled states even after a relatively small number of computation steps represent those of the full state ensemble. The cluster algorithm should however only be applied at temperatures close to the critical temperature since, as noted above, away from $T_c$ the extent of the accessible phase space and hence the intrinsic error of the single spin flip method is far smaller. [15]

## 2 Numerical Methods

Numerous algorithms can be formulated that incorporate the Wolff algorithm into transition matrix methods. As an example, a particularly simple strategy that we have implemented first generates a large ensemble of uncorrelated states of the spin system at $T_c$ with the cluster algorithm. Then for each of the states in the ensemble, the inverse temperature is first increased slowly from $\beta_c = 1/T_c$ to a certain maximum value and then, for the same state, decreased from $\beta_c$ to an appropriate minimum inverse temperature while performing single spin flips mediated by the Metropolis acceptance rule. That is, starting from a system realization "old" a single spin at a random position is reversed yielding a transition from an energy $E_{\text{old}}$ to an energy $E_{\text{new}}$. This transition is accepted with a probability given by $p_{\text{acceptance}} = \min(e^{\beta(E_{\text{old}} - E_{\text{new}})}, 1)$. At the same time, the transition matrix element corresponding to a transition from the bin associated with $E_{\text{old}}$ to the bin corresponding to $E_{\text{new}}$ is incremented by unity. After all transitions are stored, an unbiased transition matrix is obtained after normalization.

In contrast to methods such as that of the preceding paragraph, the computational efficiency can be greatly enhanced by only applying the Wolff method near the critical temperature. Accordingly, in the integrated single spin-flip/cluster flip calculation in this paper, the transition matrix is populated exclusively with single spin-flip proposals except within a region $P \equiv [\beta_c - \Delta\beta < \beta < \beta_c + \Delta\beta]$ around the inverse critical temperature. Even within this region, however, most proposals should be single-spin flips since the elements of $T$ can only be accumulated from single spin-flip proposals. In the computation below, inside the region $P$ Wolff cluster flips occur more frequently when $\beta$ is close to $\beta_c$. In detail, a triangular function $\tau(\beta(m)) = b_{\text{upper}} + (b_{\text{lower}} - b_{\text{upper}})|\beta - \beta_c|/\Delta\beta$ with $b_{\text{lower}} \approx 0$ and $b_{\text{upper}} \geq 1$ is first defined in the interval $P$. Single spin flip steps are then performed at all $m$ in $P$ unless $\mod(m, \text{ceil}(1/\tau(\beta(m)))) = 0$; that is, when $m$ modulus the smallest integer larger than $1/\tau(\beta(m))$ is zero. If this latter condition is satisfied, the single spin-flip step is replaced by a heterogeneous, compound step consisting of $N_{\text{cluster}}$ Wolff cluster flips followed by a further $N_{\text{spin}}$ standard single spin flips. The transition matrix incorporates the transitions associated with all single spin flip steps, including those that are part of the compound steps.

## 3 Results

The feasibility and accuracy of the integrated procedure will be established through a standard benchmark calculation of the specific heat of the two dimensional Ising model with zero external magnetic field, periodic boundary conditions, and a unit amplitude ferromagnetic interaction. Since the position and magnitude of the specific heat curve maximum are the quantities most affected by numerical error, the curves are quantified by specifying only the maximum value below. First a MATLAB calculation of the specific heat curve for a $32 \times 32$ spin system is performed 120 times employing the integrated spin/cluster flip procedure described in the above paragraph. Each calculation employs $6 \times 10^6$ steps of which, however a small fraction are compound steps with $N_{\text{cluster}} = 4$ and $N_{\text{spin}} = 10$. The remaining parameters are $\Delta \beta = 0.15$, $b_{\text{lower}} = 0.05$, $b_{\text{upper}} = 1.05$. Although non-optimal, to simplify the procedure an inverse temperature schedule, $\beta(m)$, is employed that varies linearly with $m$ between a starting value of $1.7\beta_c/N_{\text{taper}} \approx 0$ and a final value of $1.7\beta_c$. Forming a histogram from the maxima of the resulting 120 specific heat curves yields the result of Fig. 1, which compares with the exact result of 1.9045. [41] [42] [43] Each specific heat curve evaluation incorporated $4 \times 7.2 \times 10^5$ Wolff cluster and $12.5 \times 10^6$ single spin flips and required 104 minutes to complete on an Intel i7 processor (however 6 calculations were executed simultaneously). As expected from the relatively small number of evaluations in each specific heat calculation together with the linear temperature schedule, the results in Fig. 1 display a comparatively large statistical spread.

Three separate comparisons can be drawn between the integrated transition matrix result of Fig. 1 and benchmark calculations that employ the standard, unimproved version of the transition matrix method in which all Markov chain steps consist of single spin flips. In Fig. 2, $N_{\text{taper}} = 6 \times 10^6$ exclusive single spin flip steps are employed. This equals the total number of steps in Fig. 1 where a compound Wolff flip/single flip step is counted as a single step. The calculation time for each curve is reduced to 4.2 minutes but the accuracy is degraded by the linear temperature schedule to an extremely low level. The single spin flip calculation of Fig. 3 instead employs $N_{\text{taper}} = 12.5 \times 10^6$ so that the total number of single spin flips is now identical to the total number of single spin flips rather than the total number of steps employed to construct the transition matrix for each curve Fig. 1. While the computation time then exceeds that of Fig. 2 by a factor of two, the accuracy is not significantly improved. Finally, Fig. 4 displays the result of a single spin flip calculation with $N_{\text{taper}} = 1.2 \times 10^8$ Markov steps. This yields a 92 minute computation time per specific heat curve, roughly coinciding with that of Fig. 1. While the computational accuracy is then still noticeably less than that of the integrated procedure, the comparison is somewhat heuristic since in this calculation no attempt has been made to optimize the Wolff step schedule $\tau(\beta(m))$ given that the relative advantage of cluster algorithms in any case increases with system size.

More precisely, while the number of single spin flip steps required to obtain acceptable accuracy increases rapidly with system size, the improvement effected by a given distribution of Wolff steps is approximately size-independent. This can be easily verified by implementing a streamlined version of the above technique in which the transition matrix is populated by single spin flip transitions but a single Wolff cluster reversal is performed when the realization number $m$ falls in the region P and further satisfies $\mod\big(m, \text{ceil}(J/\tilde{\tau}(\beta(m)))\big) = 0$, where $\tilde{\tau}(\beta(m))$ is given by $\tilde{\tau}$ but with $b_{\text{upper}} = 1$ and $b_{\text{lower}} = 0$.

Considering first a $16 \times 16$ system, Fig. 5 displays the region around the maxima of 20

specific heat curves generated with the integrated procedure with $J = 40$ and a total of $1.0 \times 10^7$ Wolff and single spin flip steps of which a very small fraction of these are Wolff steps. The thick black line in this and subsequent figures indicates the exact result. The curves in Fig. 5 are noticeably better than the result of the exclusive single spin flip calculation in Fig. 6 with nearly the same computation time and hence twice as many realizations. Repeating the $J = 40$ integrated method calculation for a $32 \times 32$ system yields the result of Fig. 7 while the corresponding single spin flip result with the same computation time, Fig. 8, requires 7 times more steps. These figures clearly demonstrate that the relative improvement associated with including Wolff steps increases with system size.

The dependence of the enhancement afforded by Wolff reversals on system size is further illustrated in Figs. 9 and 10 which contain 16 specific heat curves (displayed as crosses) for a $64 \times 64$ Ising system. Fig. 9 is generated with a total of $2 \times 10^8$ realizations with $J = 800$ (hence again only a very small fraction of these are Wolff cluster reversals). In comparison, Fig. 10 displays the result of a exclusive single spin flip calculation with $3.25 \times 10^8$ realizations that requires nearly the same computation time as the calculation of Fig. 9. As expected, the improvement in the specific heat curves is comparable to that of the $32 \times 32$ spin calculation as a result of the similar number of Wolff cluster reversals.

To generate Figs. 9 and 10, we implemented a particularly simple and accurate method to calculate the density of states, $\Omega(E_i)$ from the transition matrix. For large systems repeated multiplication of an initially random vector by the transition matrix requires extended precision arithmetic because of the very large dynamic range of the density of states. A alternative procedure instead employs the detailed balance relation

$$\Omega(E_i)T_{i,i+1} = \Omega(E_{i+1})T_{i+1,i} \tag{1}$$

to recursively generate the density of states from an initially unnormalized assumed peak value. [16] [18] [22] Unfortunately, while extremely simple to implement, the accuracy of the specific heat curve is degraded compared to the repeated multiplication procedure. We therefore slightly modified the technique by instead employing the relationship [22]

$$\Omega(E_i)T_{i,i+2} = \Omega(E_{i+2})T_{i+2,i} \tag{2}$$

where the unnormalized density of states maximum is initially set to unity while its value at the histogram bin energy adjacent to the maximum is determined by Eq.(1). Because of the larger number of samples and therefore greater statistical accuracy of the second co-diagonal transition matrix elements, the specific heat curves generated by Eq.(2) were found to be nearly as accurate as those generated by the extended precision repeated multiplication method.

Finally, it should be noted that for very large numbers of realizations both the single spin flip and the integrated procedures uniformly sample all physically accessible states, as previously discussed in [28] [34] [35] [36]. Hence, the accuracy in this limit will typically only depend on the total number of realizations. In calculations, however, for which the time required for a single realization precludes attaining such a high level of precision, the combined Wolff and single spin flip technique should prove to be of great utility.

## 4  Discussion and Conclusions

Although this paper only considers the integration of local, single spin-flip and global cluster inversion methods in the context of the Ising model, numerous potential improvements and applications could be implemented. For example, one modification that we have coded exclusively applies the cluster algorithm within a narrow temperature region encompassing $T_c$ while a second generates multiple uncorrelated realizations with the cluster algorithm at $T_c$ and evolves each these separately to both high and lower temperatures while incrementally removing Markov chains as the temperature is raised or lowered. Note in this context that in vector programming languages multiple simultaneous Markov chains can achieve greatly improved computational efficiency with minimal modifications to existing code.

While for the $32 \times 32$ Ising model calculations the increase in computational efficiency afforded by the integrated algorithm, while clearly evident, is not excessive, the relative advantage of global as opposed to local spin flips will be enhanced for greater system and hence cluster sizes. Further, the integrated procedure contains several adjustable quantities such as the functional dependence of the temperature annealing schedule on the computational step number and the frequency of the cluster flips. While this paper employs a simple form for these parameters to establish that functional code can be written with a minimum of programming effort, a careful optimization strategy would presumably lead to significant efficiency increases in more involved calculations.

Finally, it should be mentioned that the cluster procedure itself has numerous implementations. Besides the standard procedures [44] [40] [45] [46], if the same calculation is performed numerous times on different systems it could be efficient to generate a database of uncorrelated samples that can then be rapidly accessed whenever a new global step is required. Alternatively, machine learning techniques that map the physical system near the critical temperature onto a more easily manipulated model have been proposed for rapidly generating global steps [38] [39] and would presumably similarly enable significant increases in computation efficiency relative to the current approach. In effect, such procedures generalize cluster algorithms to arbitrary spin models and thus significantly extend the applicability of the transition matrix cluster formalism.

## Acknowledgments

The Natural Sciences and Engineering Research Council of Canada (NSERC) is acknowledged for financial support.

# 5   Figures

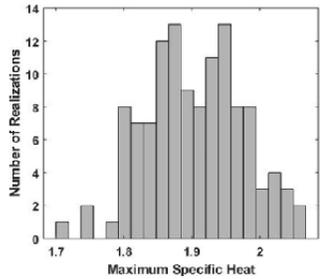

Figure 1: The histogram of the maxima of the specific heat curves obtained after 120 independent $32 \times 32$ Ising model calculations with the integrated transition matrix procedure each of which incorporated $2.8 \times 10^6$ cluster flips and $12.5 \times 10^6$ single spin flips.

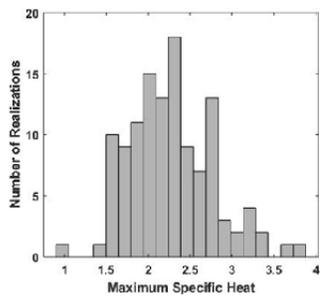

Figure 2: As in Fig. 1 but for $6 \times 10^6$ single spin flips per specific heat curve

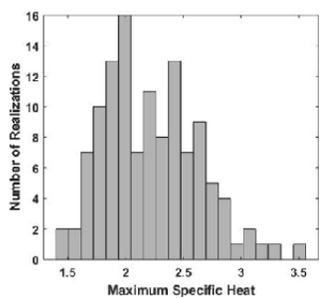

Figure 3: As in Fig. 2 but for $12.5 \times 10^6$ single spin flips.

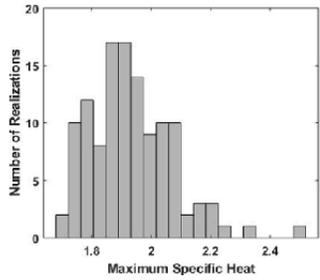

Figure 4: As in Fig. 2 but for $1.2 \times 10^8$ simultaneous single spin flips.

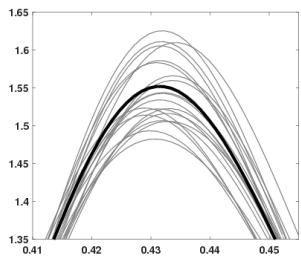

Figure 5: The region around the maxima of the specific heat curves for 20 independent $16 \times 16$ Ising model calculations with the integrated transition matrix procedure with a total of $10^7$ realizations per calculation. The thick black line in this and subsequent figures designates the exact result.

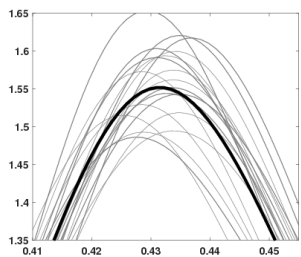

Figure 6: The results corresponding to the previous figure but for a calculation with nearly identical computation time that exclusively employs single spin flips.

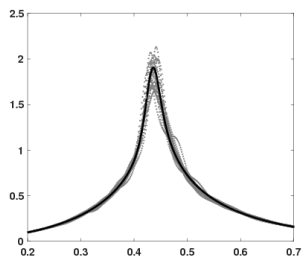

Figure 7: As in Fig. 5 but for a $32 \times 32$ spin system.

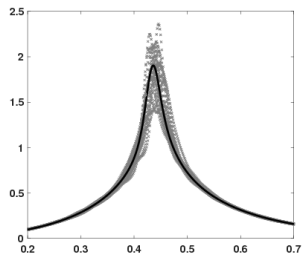
Figure 8: As in Fig. 6 but for a $32 \times 32$ spin system.

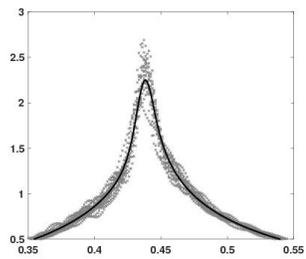
Figure 9: As in Fig. 5 but for 16 calculations of a $64 \times 64$ spin system with $2 \times 10^8$ realizations per calculation and a very small fraction of Wolff cluster reversals.

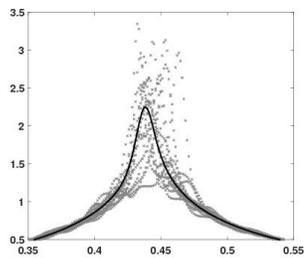
Figure 10: The results of an exclusively single spin flip calculation with a computation time nearly identical to that of Fig. 9.